\documentclass[conference]{IEEEtran}
\IEEEoverridecommandlockouts

\usepackage{cite}
\usepackage{amsmath,amssymb,amsfonts}
\usepackage{graphicx}
\usepackage{textcomp}
\usepackage{xcolor}
\def\BibTeX{{\rm B\kern-.05em{\sc i\kern-.025em b}\kern-.08em
    T\kern-.1667em\lower.7ex\hbox{E}\kern-.125emX}}

\usepackage{xspace}
\usepackage{booktabs}
\usepackage{listings}
\usepackage{algorithm}
\usepackage{algpseudocode}
\usepackage{wrapfig}
\usepackage{url}

\usepackage{physics}
\usepackage{amsmath}
\usepackage{tikz}
\usepackage{mathdots}
\usepackage{cancel}
\usepackage{color}
\usepackage{siunitx}
\usepackage{array}
\usepackage{multirow}
\usepackage{gensymb}
\usepackage{tabularx}
\usepackage{extarrows}
\usepackage{booktabs}
\usetikzlibrary{fadings}
\usetikzlibrary{patterns}
\usetikzlibrary{shadows.blur}
\usetikzlibrary{shapes}
\usepackage{enumitem}

\newcommand{\rev}[1]{#1}

\newcommand{\ms}[0]{\textsc{MemScope}\xspace}
\renewcommand{\thesubsubsection}{\Alph{subsubsection})}
\newcommand{\pld}[0]{\textsc{PL-DRAM}\xspace}
\newcommand{\dr}[0]{\textsc{DRAM}\xspace}

\usepackage{pifont}
\usepackage{colortbl}
\newcommand{\Yes}{\ding{51}}%
\newcommand{\No}{\ding{55}}%
\newcommand{\niparagraph}[1]{\noindent\textbf{#1}\hspace{0.5em}}
\begin{document}

\author{
    Golsana Ghaemi\IEEEauthorrefmark{1},
    Gabriel Franco\IEEEauthorrefmark{1},
    Kazem Taram\IEEEauthorrefmark{2},
    Renato Mancuso \IEEEauthorrefmark{1}\\
    \IEEEauthorrefmark{1}Boston University\\
    \IEEEauthorrefmark{2}Purdue University \\
    Email: \{golosana,gvfranco, rmancuso\}@bu.edu, kazem@purdue.edu
}

\title{Heterogeneous Memory Benchmarking Toolkit}

\maketitle

\begin{abstract}
This paper presents an open-source kernel-level heterogeneous memory characterization framework (\ms) for embedded systems. \ms enables precise characterization of the temporal behavior of available memory modules under configurable contention stress scenarios. \ms leverages kernel-level control over physical memory allocation, cache maintenance, CPU state, interrupts, and I/O device activity to accurately benchmark heterogeneous memory subsystems. This gives us the privilege to directly map pieces of contiguous physical memory and instantiate allocators, allowing us to finely control cores to create and eliminate interference. Additionally, we can minimize noise and interruptions, guaranteeing more consistent and precise results compared to equivalent user-space solutions. Running our Framework on a Xilinx Zynq UltraScale+ ZCU102 CPU-FPGA platform demonstrates its capability to 
precisely benchmark bandwidth and latency across various memory types, including PL-side DRAM and BRAM, in a multi-core system. 


\end{abstract}


\section{Introduction}\label{sec:intro}

The ever-increasing demand for high-performance systems, combined with the steady rise in data-intensive processing workloads, has been a defining force for the modern landscape of hardware platforms. The push for higher performance has impacted general-purpose systems and embedded/real-time systems. System \emph{heterogeneity} has been pivotal in the last decade of embedded systems evolution [embedded\_heter] and the subject of a plethora of studies~\cite{survey_heter}.

Modern high-performance systems-on-a-chip (SoCs) are characterized by high \textbf{compute heterogeneity}. Indeed, they consist of a wide range of cross-vendor computing blocks ranging from general-purpose processors (CPUs) to special-purpose accelerators and even FPGAs. Established OS-level methodologies have emerged to benchmark and support application development in heterogeneous systems. Notable examples include the Linux Remote Processor Framework~\cite{linux-remoteproc} and the OpenMP Framework~\cite{OpenMP}.

The heterogeneity in modern platforms is not limited to computing resources. \textbf{Memory heterogeneity} has co-evolved with compute heterogeneity. Different memory technologies coexist, each with specific characteristics in terms of size, cost, and temporal behavior. Not only does the baseline performance (e.g., single-threaded accesses) of these memories range widely, but so does their temporal behavior under stress (e.g., multi-threaded accesses). Notable examples of memory technologies with widely ranging characteristics include Double Data Rate (DDR), Reduced-Latency DRAM (RL-DRAM)~\cite{RLdram_hassan}, High-Bandwidth Memory (HBM)~\cite{HBM_ECRTS}, Non-Volatile Random Access Memory (NVRAM)~\cite{NVM_rt}, on-chip Static Random Access Memory (SRAM)~\cite{bus-aware, spmwasly}.

\niparagraph{Challenges.} We focus on memory heterogeneity. While heterogeneous memory subsystems present vast opportunities to optimize memory allocation for real-time and embedded applications, their practical use presents several challenges. Said challenges can be grouped into \emph{Characterization Challenges} and \emph{Usage Challenges}. Characterization challenges hinder the construction and deployment of precise, controlled, and interference-free experiments to understand the temporal behavior of memory modules when relying on conventional user-space toolkits. Usage challenges prevent the efficient allocation of heterogeneous memory to user-space applications.

\niparagraph{\ms as the Proposed Solution.} In this paper, we address characterization challenges. To do so, we design, implement, and evaluate a novel open-source in-kernel heterogeneous memory characterization toolkit called \ms. \ms is designed as a Linux kernel module, requiring no kernel source modifications, to boost broad adoption. It is designed to (1) automatically recognize heterogeneous memory modules described via the kernel device tree; (2) internally instantiate per-memory allocators under the direct control of system evaluators; (3) provide an extensible library of micro-benchmarking activities; (4) allow intuitive experiment definition and results retrieval from user-space; and (5) minimize experimental noise with direct control over CPU and interrupt state during an active experiment. 

\niparagraph{Contribution.} This paper makes the following contributions. (1) We propose the first kernel-level heterogeneous memory characterization framework, namely \ms; (2) We provide a full open-source implementation of \ms; (3) We evaluate the capabilities of \ms on a modern embedded platform featuring a high degree of memory heterogeneity; (4) We demonstrate that \ms allows accurate characterization with valuable insights to drive memory allocation in user-space applications.

\section{Motivation and Goal}\label{sec:motivation}
Attention to memory management in heterogeneous systems has received substantial interest from the general-purpose and high-performance systems computing community, as we review in Section~\ref{sec:relwork}. 

Nonetheless, no \emph{de facto} turnkey solution exists to perform heterogeneous memory characterization efficiently. \ms aims to fill this gap, primarily targeting Linux-based high-performance real-time embedded systems.

\begin{figure*}[t]
    \centering
    \includegraphics[width=0.8\linewidth]{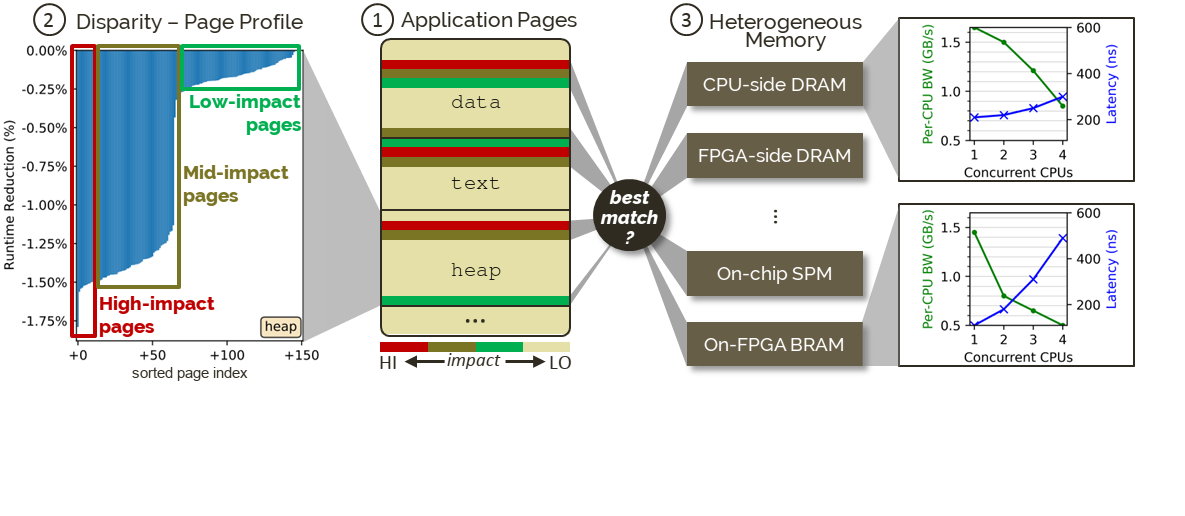}
    \vspace{-2cm}
    \caption{The problem of heterogeneous memory management consists in performing memory allocation given (1) proper characterization of the temporal behavior of memory modules due to technological heterogeneity (right-hand side), and (2) the expected impact of a given memory page on the temporal behavior of applications due to usage heterogeneity (left-hand side).}
    \label{fig:heter_mgmt}
    \vspace{-0.5cm}
\end{figure*}

\subsection{Sources of Memory Heterogeneity}
Heterogeneous memory subsystems amplify the complexity of proper management for time-sensitive applications due to the interplay of two effects, namely \emph{technological heterogeneity} and \emph{usage heterogeneity}, as depicted in Figure~\ref{fig:heter_mgmt}.

\niparagraph{Technological Heterogeneity.} As briefly mentioned in Section~\ref{sec:intro}, memory modules differ in size, cost, and inherent temporal characteristics, such as read/write latency and bandwidth. Several hardware-level characteristics contribute to the exhibited temporal characteristics, such as (1) the type of memory cells they comprise (SDRAM, SRAM, or NVRAM) which impacts their performance, power, and persistence characteristics; (2) their architectural organization---e.g., SDRAM cells can be flatly arranged to in traditional DRAM systems, or 3D-stacked in High-bandwidth Memory (HBM) modules; SRAM cells can be used to define architectural caches, scratchpads, or in FPGAs as Block RAM (BRAM) and ultraBRAM modules.

Moreover, different memory types exhibit varying performance characteristics under contention, owing to their intrinsic memory-level parallelism (MLP). As such, bandwidth and latency can be impacted by interference from concurrent tasks or competing memory requests from multiple cores, leading to nonlinear performance degradation. 
As depicted on the right-hand side of Figure~\ref{fig:heter_mgmt}, memory modules in a heterogeneous memory subsystem are characterized by \textbf{performance curves} parametrized by the type of accesses and the degree of contention. For instance, traditional CPU-side DRAM might exhibit worse single-threaded latencies than an FPGA-side scratchpad (BRAM) but sustain better multi-threaded bandwidth as concurrent accesses increase.

\niparagraph{Usage Heterogeneity.} Memory resources are often the performance bottleneck in data-heavy workloads. Depending on the application, low-latency access to some memory pages might largely impact the execution time. Conversely, placing other pages in slow memory might have a negligible impact. Fortunately, the need to profile the demand of applications for memory resources is well understood~\cite{Gleipnir_13, Alleria_2019, bbprof_ecrts21}. Borrowing and annotating a figure from~\cite{bbprof_ecrts21}, the left-hand side of Figure~\ref{fig:heter_mgmt} depicts the per-page runtime reduction percentage when individual heap pages are allocated in cache.
\subsection{Key Challenges}

Inspired by the famous quote \emph{``You can't manage what you don't measure,"} often attributed to Peter Drucker, we aim to systematically analyze the temporal characteristics of heterogeneous memory subsystems in embedded systems, gather deeper insights into performance variations and optimize memory usage.

To this end, propose an extensible and easy-to-use kernel-based benchmarking infrastructure addressing the key challenges (\textbf{C1}--\textbf{C5}) reviewed below.

\niparagraph{C1: Imprecise Physical Memory Allocation.}
In user-space, memory allocation is mediated by the virtual memory layer. Thus, limited control can be exerted over physical memory allocation. This shortcoming poses a fundamental challenge when characterizing heterogeneous memory.

\niparagraph{C2: Imprecise Compute Engine Activity.}
To evaluate memory performance under isolated conditions, one must control the execution context of the benchmarking activities. In user-space, it is challenging to prevent system daemons, kernel threads, and background processes from interfering.

\niparagraph{C3: Imprecise Interrupt Activity.} In user-space, applications cannot disable or redirect interrupts, nor can they prevent the kernel or other subsystems from servicing them on the core of interest. This leads to two major sources of noise: (1) interrupts can preempt benchmarking tasks, and (2) servicing interrupts may generate additional memory traffic.

\niparagraph{C4: Restricted Cache Maintenance.} Caches often act as an opaque layer that masks the true behavior of the underlying memory. Thus, controlling cache states is key to accurately assessing memory performance. User-space applications, however, are restricted in their access to cache maintenance instructions or cache-control interfaces.

\niparagraph{C5: Restricted Access to Performance Counters.} Hardware performance counters offer fine-grained visibility into metrics that are invaluable when dissecting the behavior of complex memory systems. Unfortunately, user-space access to performance counters is often limited or highly abstracted.

\subsection{The \ms Approach}

To attain full control over allocation strategies, access patterns, cache invalidation, access to performance monitors, and CPU states, we implement our benchmarking infrastructure at the kernel level. A similar motivation fueled the seminal work on NanoBench~\cite{nanobench}, the only kernel-level toolkit for single-core CPU benchmarking. \ms is the first kernel-level toolkit for the characterization of heterogeneous memory subsystems in multicore systems.

On top of what was mentioned above, due to the widely varying temporal characteristics of different memories, gaining insight into these variations allows us to understand how application runtimes are impacted by allocation decisions, especially crucial for safety-critical real-time embedded systems. Certain pages will experience more or less interference depending on allocation strategies, affecting overall performance. By analyzing these behaviors, we can make informed decisions about memory allocation to mitigate contention and optimize execution. Our work opens the door for future integration into memory allocation, utilizing heterogeneous memories. In the next section ~\ref{sec:design}, we will outline the blueprint of our approach, explain the design challenges, and how we tackle these challenges.

\begin{figure*}
    \centering
    \includegraphics[width=.8\linewidth]{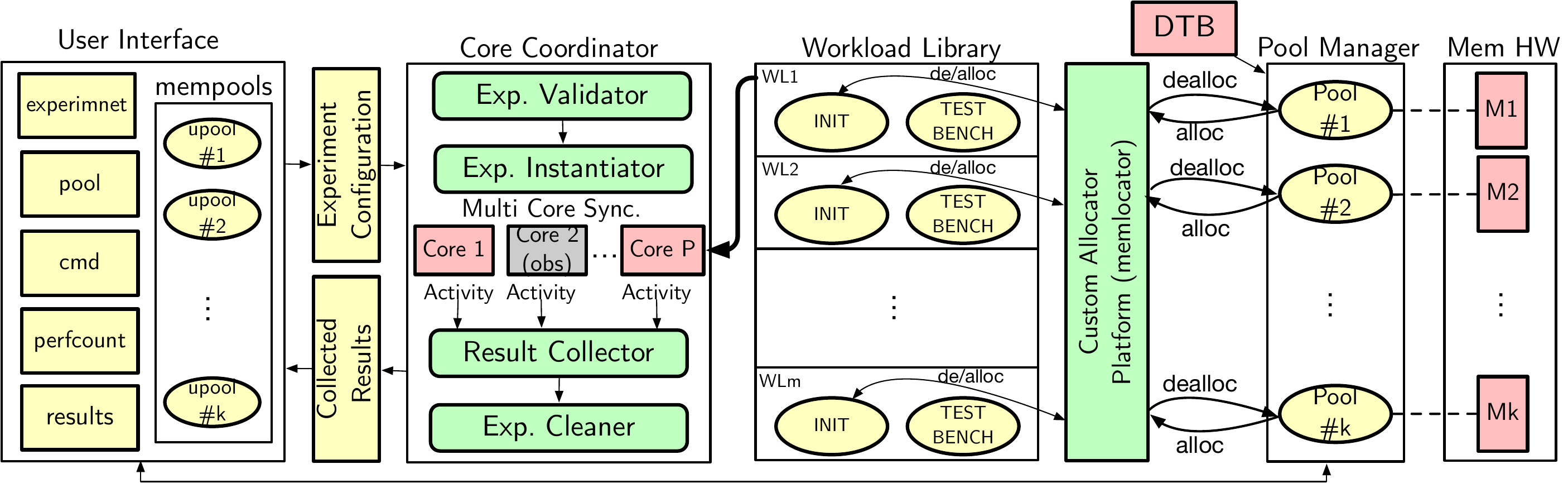}
    \caption{High-level structure of \ms highlighting its main components and their interplay.}
    \label{fig:interplay}
    \vspace{-0.5cm}    
\end{figure*}

\section{\ms Design}\label{sec:design}

In this section, we describe the primary design elements of \ms. The overall system design, depicted in Figure~\ref{fig:interplay}, comprises four main components. First, we cover the structure of each benchmarking experiment in \ms---see Section~\ref{sec:exp_structure}. Next, we discuss the various sub-modules depicted in Figure~\ref{fig:interplay} that are crucial for the following functionalities: (1) memory target selection via a \emph{Memory Pool Manager} (Section~\ref{sec:pm}); (2) access pattern selection via the \emph{Workload Library} (Section~\ref{sec: wl}); (3) multi-CPU orchestration via the \emph{Core Coordinator} (Section~\ref{sec:cc}); (4) user interaction for experiment control and result retrieval (Section~\ref{sec:ui}).

\ms is open-source and the full code is available in the project repository\footnote{Repository link omitted to comply with double-blind requirements. Repository link will be disclosed to the PC chair.}. For the sake of conciseness, we defer the reader interested in the low-level implementation details to the supplementary material and keep the discussion in this section focused on the high-level design principles.

\subsection{Experiment Structure in \ms}\label{sec:exp_structure}
The goal of a \ms experiment is to evaluate the temporal characteristics of a target memory module under a varying degree of contention generated by the other online CPUs. As such, each experiment in \ms consists of a sequence of \emph{scenarios}. Each scenario is comprised of a set of \emph{monitored activities} across all online CPUs. All experiments follow a common structure:

\begin{enumerate}[wide, labelwidth=!, labelindent=0pt]
    \item Memory targets and access pattern parameters for the core under observation and stressor cores are runtime configurable.
    \item The temporal behavior of the observed core is measured following a sequence of increasingly worse stress scenarios.
    \item Scenario-specific workloads are assigned to both the core under observation and all the interfering cores. The workload assigned to the core under analysis can differ from the one executed by a stressor core.
    \item Micro-architectural events are collected for all CPUs.
    \item Results include the total bytes read/written from/to the target memory, the execution time for the core under observation, and the sampled architectural events across all cores.
  
    \item At the end of each scenario, and also upon the completion of the entire experiment, \ms performs per-core data structure management and deallocates all allocated buffers to ensure a clean state for subsequent experiments.
\end{enumerate}
Scenarios, ranging from the \emph{best} to \emph{worst} case, are executed in an automated sequence. In the best scenario, the core under observation runs the selected workload while all other cores remain \emph{memory-idle} by executing a CPU-intensive, non-memory workload. Once this scenario completes and the results are collected, the second scenario begins: one additional core starts executing the stress workload while the rest remain memory-idle. In the following scenario, \ms increases the number of stress cores by one. This process continues until the worst-stress scenario: all available cores are actively stressing the selected target memory.

\subsection{Memory Pool Manager}\label{sec:pm}
A benchmarking infrastructure for heterogeneous memory subsystems requires designing mechanisms to precisely select the target memory pools. 

To this end, \ms leverages the same mechanisms that the OS uses to describe hardware resources, i.e., \emph{device trees}, to auto-detect an arbitrary number of available memory areas. 

\ms instantiates a set of Linux kernel-compatible \emph{memory pools}, one per detected memory module, leveraging the \texttt{genalloc/genpool} kernel subsystem. Thanks to the 1-to-1 correspondence between memory pool IDs and hardware memory modules, \ms allows to select memory targets via allocation pool IDs. The example presented in Figure~\ref{fig:interplay} depicts the memory pool manager and the creation of pools with IDs \#$1$ to 
\#$k$ from available underlying memory modules \mbox{M$_1$, M$_2$, \ldots, M$_k$}.

In our evaluation setup, for instance, we instantiated memory pools from multiple memory technologies present in our setup, including DRAM, FPGA-side DRAM (PL-DRAM), FPGA-side Block RAM (BRAM), and On-Chip Memory (OCM). The pool manager eliminates the need for manual detection and configuration of memory pools parameters, enhancing flexibility. This design also allows for the seamless integration of additional memory technologies, e.g., Non-Volatile Memory (NVM) and disaggregated remote memory.

The instantiated memory pools are primarily used internally to conduct memory performance experiments. In addition, \ms also exports these pools for memory allocation in user space (\emph{upools}). It does so by extending the user interface and  creating a set of device files aptly named \texttt{/dev/upool<ID>} that can be memory-mapped by applications to allocate pages from the corresponding pools. Details of this part are available in the supplementary material.

\subsection{Workload Library}\label{sec: wl}
Apart from selecting the memory to benchmark, \ms also allows one to select the performance metric to be measured for the chosen memory target. The specific choice depends on the particular features of the memory subsystem one wish to analyze, as well as the stress/memory contention scenarios for which insights are desired. It is important to note that depending on the experiment parameters, \ms allows to benchmark not only the target memory module, but also its interplay with CPU caches and bus architecture, as demonstrated in our evaluations---see Section~\ref{sec:evaluation}.

To this end, the workload library offers a suite of configurable micro-benchmarking workloads, each designed to shed light on a set of specific performance parameters. As such, the included test benches are registered in the library based on the access patterns they implement. 

This modular approach ensures flexibility and ease of maintenance when expanding or modifying the workload library.

\niparagraph{Configurable Buffer Initialization.} \ms allows the definition of a per-workload buffer initialization routine. This is invoked before activating the corresponding workload to initialize the target memory buffer as needed.

\niparagraph{Access Strategies.}
Our library currently focuses on bandwidth and latency measurements of memories under various access strategies, including (1)~normal read, (2)~normal write, (3)~non-cacheable read, (4)~non-cacheable write, (5)~non-cacheable write streaming, and (6)~read/write with non-temporal load/store instructions. Non-cacheable operations refer to access strategies that bypass the CPU caches, ensuring that read/write operations directly interact with the target memory. These allow measuring the performance of memory modules (e.g., scratchpad) that are smaller than the last-level cache. Finally, non-temporal access patterns are implemented through architectural features that allow specific load/store operations to bypass caches.

\niparagraph{Bandwidth Measurement Workloads.} The goal of the bandwidth micro-benchmarks included in \ms is to estimate the throughput that a target memory module is capable of sustaining at steady state. Since the goal is to maximize the rate of transactions generated by the core and to avoid compiler effects, all our bandwidth measurement test benches are directly implemented in assembly. These micro-benchmarks perform sequential accesses to the provided buffer at the cache line granularity.

\niparagraph{Latency Measurement Workloads.}
The goal of these workloads is to compute the average round-trip time for a generic memory request. To ensure precise measurements, these workloads must ensure that only one outstanding memory operation at a time is emitted by the core under analysis. To do so, we leverage data dependencies. Thus, we ensure that the next memory location to be accessed is only known once the data for the previous access has been completed. We devise an approach that ensures full coverage of the target buffer while remaining impossible to prefetch. The details are provided in Appedix~\ref{sec:impl} (see supplementary materials).

\niparagraph{Memory-Idle Workload.} In addition to all the mentioned \emph{memory-bound} workloads, a "busy loop" test bench is included for \emph{memory-idle} benchmarking. The busy CPU-bound loop, in combination with strict kernel preemption and interrupt control, allows us to keep the core inactive in memory.

\subsection{Core Coordinator}\label{sec:cc}
When an experiment is launched, the core coordinator is responsible for (1)~validating the experiment configuration, (2)~deploying all the workloads, (3)~managing the synchronization between the cores, and (4)~aggregating the final results. Thus, \ms's core coordinator includes two primary components, namely the \emph{Experiment Instantiator} and the \emph{Multi-core Synchronizer}.

\niparagraph{Experiment Instantiator.} Once the experiment configuration has been received, the instantiator is invoked. It checks the sanity of the experiment parameters, such as the buffer size, access type, availability of pages in the selected pool, etc. If validation passes, the instantiator spawns the scenario-specific workloads on the online cores. These are referred to as \emph{activities} while they are dispatched on the cores.

There are three main groups of activities that must be managed. First, the \emph{Main Activity} runs on the core under observation and can be any benchmark from the workload library. Next, the \emph{Stress Activity} corresponds to the workload active on a stressor core. Once again, the nature of this activity can be selected from the workload library. Finally, the \emph{Idle Activity} runs on all the cores that must remain memory-idle during the current scenario. In this case, the busy-loop workload is automatically selected.

Besides managing buffers and data structures related to activities, \ms samples the selected performance counters for all cores. Configuring, enabling, and later disabling these performance counters based on the user parameters for each core is another crucial responsibility.

After spawning the appropriate activities and enabling the selected performance counters, the next key responsibility of the core coordinator is managing synchronization between cores during the execution of activities, which is taken care of by the next submodule.

\niparagraph{Multi-Core Synchronization.}
With $p$ online CPUs, \ms measures the temporal behavior of the target memory and with the selected stress workload across $p$ scenarios, as described in Section~\ref{sec:exp_structure}. A key challenge is ensuring that all the idle/stressor cores have truly initiated the current activity before any measurement on the observed core is performed. If this was not the case, the obtained measurements might capture a partial overlap between the observed core's activity and the other cores not appropriately stressing the target memory or still acting as stressors as part of a past activity instead despite being expected to be idle. This situation might lead to inaccurate results and non-repeatable results. 

Similarly, stopping activities requires synchronization. The core coordinator cannot simply issue a stop command and proceed to the next run without verifying that all the other cores have actually ceased execution. Due to potential delays in processing stop commands, an immediate transition could result in overlapping execution between scenarios, once again impacting measurement validity.

To ensure accuracy and repeatability, we enforce the following constraints: (1) Measurement on the observed core begins only after all stressor/idle cores have started activity execution; (2) The experimental scenario remains stable throughout the measurement period; (3) Measurement on the observed core stops before any stop command is issued for the other cores; and (4) The next scenario does not begin until all stressor/idle cores have fully completed execution of the previous run. These constraints are strictly enforced by leveraging kernel-level synchronization mechanisms, as detailed in Appendix~\ref{sec:impl}.

\subsection{User-Space Interface}\label{sec:ui}
The user-space interface module serves as the primary entry point for interaction with \ms to configure, launch experiments, and retrieve results. For this purpose, it exposes a number of entries briefly reviewed below.

\niparagraph{Experiment  Configuration Entry.}
Each experiment requires multiple parameters to be configured. This entry accepts a configuration string where said parameters can be specified in a positional manner. These include (1) memory mapping type---e.g., normal cacheable, strongly ordered, shareable, and so on\footnote{Due to the already large number of parameters, in this paper we only consider normal cacheable memory mappings.}; (2) memory access pattern---e.g., sequential for read/write bandwidth measurement, with dependencies for latency measurements, sequential but non-cacheable, write-streaming, etc.; (3) buffer size to allocate and access; (4) target memory pool. Two sets of parameters (1)--(4) must be specified, one for the core under observation and one that will be used for all the stressor cores.

\niparagraph{Performance Counter Selection.}
\ms is designed to use all the available performance counters during an experiment. This entry allows the selection of two sets of performance events to be monitored with the available hardware performance counters. The first set will be configured on the core under observation, while the other set will be used on all the idle/stressor cores.

\niparagraph{Pools Status.}
Through this entry, one can retrieve the full list of available memory pools as detected by \ms at load time. For each pool, this entry reports the pool ID, the corresponding size, the physical address mapping base, and the number of pages available for allocation.

\niparagraph{Results.}
This entry allows access to the collected results in a user-readable format. The results include the main temporal measurements, the amount of memory read/written during the experiment, and the final value of the considered performance counters. The entry also reports the configuration setup used to gather the results.

\niparagraph{Experiment Command.}
Finally, this entry enables experiment control. Once an experiment is configured, it can be launched via a \emph{start} command. It is also possible to trigger experiment \emph{validation} without launching the configured experiment. Finally, the result from the previous experiment can be \emph{erased}, freeing the associated resources.

\section{Evaluation} \label{sec:evaluation}
In this section, we present our evaluation of \ms, starting with our methodology and platform setup. The rest of the section presents four classes of experiments:
(1)~Characterization of DRAM variants and their performance under contended access.
(2)~Benchmarking of on‑chip scratchpad memories to assess their temporal behavior.
(3)~Analysis of cache microarchitectural behavior and the impact of cache partitioning. (4)~Studying the impact of heterogeneous memory management on real-world applications.

\begin{figure}
  \centering
  \includegraphics[width=0.8\linewidth]{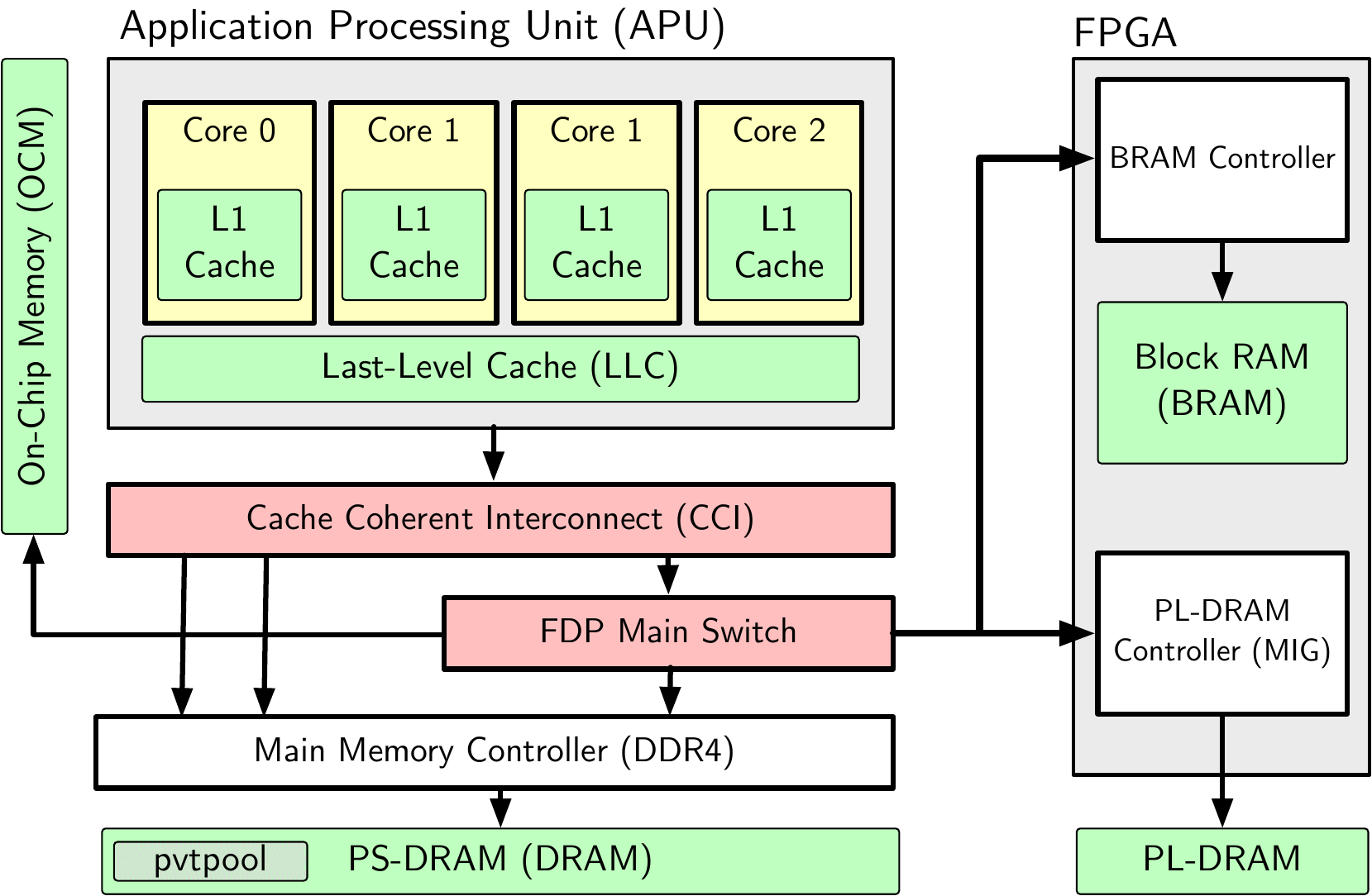}
  \vspace{-0.2cm}
  \caption{Overview of evaluation SoC including heterogeneous memory pools.}
  \label{fig:zcu}
  \vspace{-0.5cm}
\end{figure}

\subsection{Experimental Methodology} \label{sec:eval_meth}

\ms was implemented and tested on Linux kernel v5.4 and evaluated on a Xilinx-ZCU102 development platform featuring a Zynq UltraScale+ XCZU9EG MPSoC~\cite{xil_zcu102}, depicted in Figure~\ref{fig:zcu}.
The main processor is a 64-bit quad-core ARM Cortex-A53~\cite{cortex_a53} which uses ARMv8-A~\cite{arm_v8a} ISA and operates at 1.5 GHz.
L1 cache comprises 32KB/64KB instruction/data cache with 2-way/4-way set-associativity. The last level cache (L2) is a unified 16-way set associative cache with size of 1MB. The LLC is shared among all cores. The cache line size is 64 bytes for both cache levels.

As shown in Figure~\ref{fig:zcu}, our platform features 4 types of memories:
(1)~the DRAM module that is directly connected to the CPU cluster (\emph{PS-DRAM}), which we refer to as \emph{\dr}; (2)~the DRAM module that is connected to the programmable logic (\emph{\pld}) (3) on-chip scratchpad memory (\emph{OCM}) and, (4) the FPGA-side block random access memory (\emph{BRAM}). In our platform’s Device Tree Blob (DTB), we expose the following memory regions for MemScope’s allocator:
128\,KB of OCM, 1\,MB OF BRAM, 256\,MB of \dr and \pld. These sizes represent the slices we carve out for benchmarking; the underlying hardware supports larger capacities.

For the experiments in Section~\ref{sec:cacheanalysis}, we use cache partitioning via page coloring through the Minerva Jailhouse~\cite{minerva_jh}.

Cache partitioning allows us to isolate the effects of conflicts on cache sets from the effects of contention on downstream memory modules and shared bus segments.

This configuration defines two contiguous intermediate physical address (IPA) ranges. The first IPA range includes all normal memory used by Linux and is mapped by Jailhouse to 12 out of 16 (i.e., 3/4) of the available colors. The second range is mapped to pages using the remaining 4 out of 16 (i.e., 1/4) colors. This range is then exported to \ms as a memory pool. As such, only benchmarks with pages allocated from this pool will be able to use the \emph{private} 25\% portion of the L2 cache (256 KB). We refer to this pool as the \emph{private cache pool}, namely \emph{pvtpool} in our experiments. From \ms's point of view, this is a distinct heterogeneous memory module.

\ms supports configurable iteration counts for workload execution to ensure the statistical stability of the measured performance metrics. In all experiments discussed in this section, we configured this iteration count to $500$.

\newcommand{\mpat}[1]{\texttt{(#1)}\xspace}

In our results, we use different access strategies. The full list of supported strategies is provided in Table~\ref{tab:accesspattern}.
We use tuples of the form \mpat{a,b}, where \texttt{a} indicates the access strategy employed by the core under observation while \texttt{b} that of a stressor core. For instance, with \mpat{r,w}, the core under observation performs sequential reads while stressor cores execute sequential writes.

\begin{table}[t]

    \centering
    \caption{Available access strategies in \ms.}\label{tab:accesspattern}
    \resizebox{\linewidth}{!}{
    \begin{tabular}{ll}
        \toprule
        Access Pattern & Description \\\midrule
        \texttt{r} & sequential reads to benchmark memory read bandwidth \\
        \texttt{w} & sequential writes to benchmark memory write bandwidth \\
        \texttt{l} & data-dependent random reads (pointer chasing) to benchmark latency \\
        \texttt{s} & non-cachable version of the \texttt{r} benchmark\\
        \texttt{x} & non-cachable version of the \texttt{w} benchmark\\
        \texttt{m} & non-cachable version of the \texttt{l} benchmark  \\
        \texttt{y} & non-cacheable write-streaming to the memory (no write-allocate)\\
    
        \bottomrule
    \end{tabular}}
\end{table}

\subsection{Analysis of DRAM Modules} \label{sec:dramanalysis}

\begin{figure}
\centering
\includegraphics{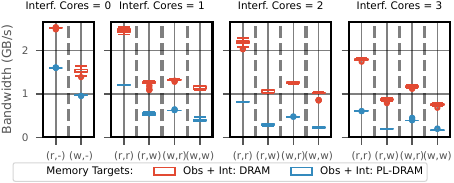}
\vspace{-0.2cm}
\caption{Homogeneous bandwidth results for DRAM and \pld under four stress scenarios with buffer size of 4 MB.}
\label{fig:dram-mig-bwboxplot}
\vspace{-0.5cm}
\end{figure}

In this subsection, we describe the results of \ms's characterization of the two DRAM memory types in our platform (\dr and \pld), in terms of bandwidth, latency, and memory-level parallelism under various scenarios.

We use \ms to understand how \dr and \pld behavior changes in isolation as operations vary and how they react under different levels of stress. To this end, we test two \emph{homogeneous} setups and two \emph{heterogenous} setups. In the homogeneous setups---Subsections~\ref{sec:dramanalysis}(1),~\ref{sec:dramanalysis}(2), and~\ref{sec:dramanalysis}(3)---we observe the behavior of the \dr (resp., \pld) while the stressors also target the \dr (resp., \pld) module. Conversely, in the heterogeneous setups---Section~\ref{sec:dramanalysis}(4)---we observe the behavior of the \dr (resp., \pld) while the stressors target the \pld (resp., \dr) module. In these experiments, the buffer size is 4~MB unless otherwise specified.

\subsubsection{\textbf{Homogeneous Bandwidth Analysis}}\label{sec:homodrambw}

Figure~\ref{fig:dram-mig-bwboxplot} shows the bandwidth extracted by the observed core from the two DRAM memory types.
As expected, it decreases as the number of interfering cores increases. However, this drop is more noticeable in \dr than in \pld. 

The \dr bandwidth drop becomes noticeable with more than one interfering core in the \mpat{r,r} case; it is substantial in the \mpat{r,w} case, even with only one stressor core. This is expected, as the cache system follows the write-allocate/write-back (WAWB) policy, meaning that every store resulting in a write miss causes both a memory read and a write-back of some dirty line being evicted. This implicit read in case of write miss can further exacerbate contention effects. 

Additionally, read operations on the core under observation are synchronous (due to its in-order nature). Thus, pending loads cause pipeline stalls that directly affect the end-to-end execution time and that are amplified if the stressors produce read+write traffic caused by store-heavy access. Conversely, for the \dr under \mpat{w,r} operations, the bandwidth remains relatively stable due to the opposite effect of the logic discussed. \pld follows a similar trend, albeit remaining consistently at a lower performance level and with proportionally lower performance degradation. The trend similarity, moreover, highlights how the behavior is characteristic of DRAM technology in spite of substantial differences in clocking, capacity, and manufacturers.

\ms allows us to make the following observations: (1)~the bandwidth of \pld is lower than \dr, as expected, due to its greater distance from the cores and lower clock domain (PL), 
(2)~scenarios exist where a stressed \dr---e.g., in the \mpat{r,w} case---exhibits a bandwidth comparable to that of a non-stressed \pld, and (3)~the stress-induced bandwidth degradation for the \dr module is proportionally more pronounced compared to \pld. A heterogeneous memory allocator can leverage these insights.

\begin{figure}

\centering
\includegraphics[width=\linewidth]{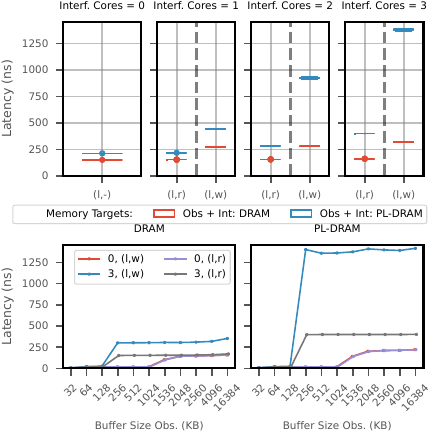}

\caption{Homogeneous latency results for DRAM and \pld under four stress scenarios with buffer size of 4 MB.}
\label{fig:dram-l-mig-l-boxline}
\vspace{-0.3cm}
\end{figure}

\subsubsection{\textbf{Homogeneous Latency Analysis}}\label{sec:homodramlat}
Figure~\ref{fig:dram-l-mig-l-boxline} shows the results of latency analysis using \ms using access strategies \mpat{l,r} and \mpat{l,w}---see Table~\ref{tab:accesspattern}. 

With increasing stress, the latency gap between best- and worst-case scenarios grows. Interestingly, \dr and \pld both start from almost the same latency. However, this gap widens as contention worsens. The change in latency for both read and write stress in \dr remains relatively stable. In contrast, \pld reacts significantly to the increase in stressors. 

The line plots at the bottom of Figure~\ref{fig:dram-l-mig-l-boxline} show the measured latency in scenarios with 0 and 3 stressor cores for increasing buffer sizes. In the 0-stressors case, caching effects disappear for buffer sizes above 1~MB; in the 3-stressors case, they disappear for sizes above 256~KB. For \dr, the latency variation from the best- to the worst-case remains stable at around of $0.3~ns$, whereas for \pld, the latency fluctuates between $1.3$ and $1.4~ns$.

\subsubsection{\textbf{MLP Derivation}}\label{sec:mlpdram}

\begin{figure}

\includegraphics{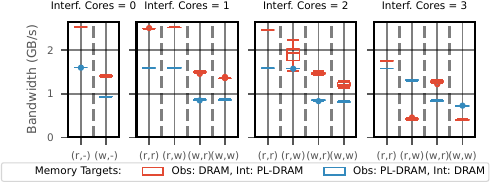}
\caption{Heterogeneous bandwidth results for DRAM and \pld under four stress scenarios with buffer size of 4 MB.}
\label{fig:dm-md-bwboxplot}
\vspace{-0.5cm}
\end{figure}

\begin{table}[h!]
\centering
\begin{minipage}[b]{0.45\textwidth}
 {\scriptsize\caption{MLP calculation for \dr}\label{tab:mlpdram}}

\resizebox{\textwidth}{!}{ 
\begin{tabular}{l l c c c}
\toprule
Lat. Operation & BW Operation & Lat.(ns/$T_x$) & BW($T_x$/ns) & MLP \\
\midrule
(l,r) & (r,r) & 161.89 & 0.03 & 4.85 \\

(l,w) & (r,w) & 318.56 & 0.014 & 4.45 \\
\bottomrule
\end{tabular}

}

\end{minipage}
\hfill
\begin{minipage}[b]{0.45\textwidth}
{\scriptsize\caption{MLP calculation for \pld}\label{tab:mlppldram}}

\resizebox{\textwidth}{!}{ 
\begin{tabular}{l l c c c}
\toprule
Lat. Operation & BW Operation & Lat.(ns/$T_x$) & BW($T_x$/ns) & MLP \\
\midrule
(l,r) & (r,r) & 399.49 & 0.01 & 3.99\\

(l,w) & (r,w) & 1386.80 & 0.003 & 4.16 \\
\bottomrule
\end{tabular}
}

\end{minipage}
\end{table}

Table \ref{tab:mlpdram} and \ref{tab:mlppldram}, display the measured \emph{Memory-Level Parallelism} (\emph{MLP}) for \dr and \pld. MLP is calculated for both memory types using \emph{Little's Law}, stating that for a system at steady state, the average MLP can be estimated as: $\text{Avg. MLP} = {\text{Avg. Latency}} \times {\text{Avg. Bandwidth}}$.

For this analysis, we use the results captured in the worst-case scenarios, where all the interfering cores are executing memory-intensive read/write operations.
For bandwidth measurements, we select cases that maximize the throughput, 
 
We evaluate the MLP perceived by the core under analysis (access strategy \texttt{l}), in the case when the other cores perform sequential reads (\texttt{r}) or writes (\texttt{w}). Thus, we pair latency experiments \mpat{l,r} with bandwidth experiments \mpat{r,r} and \mpat{l,w} with \mpat{r,w}.

We observe comparable values of MLP between the two memory modules in spite of the substantially higher latencies observed under stress for \pld. This potentially highlights that the bottleneck on the number of outstanding memory transactions lies in the bus infrastructure (CCI, see Figure~\ref{fig:zcu}) that is common for transactions targeting either module. However, because \pld transactions have significantly higher latency, outstanding transactions generally occupy bus-level queue entries for longer. This can reduce the opportunity for transactions targeting faster memory (e.g., \dr) to progress, effectively throttling its throughput.

This observation motivated the next set of experiments presented in the following section. Our goal is to investigate how mixed access to two memory systems with significant latency disparity respond under stress.

\subsubsection{\textbf{Heterogeneous Bandwidth Analysis}}\label{sec:heterdrambw}

We present a heterogeneous bandwidth and latency analysis to address the following question: \emph{how does temporal behavior change when the target memory for the core under observation differs from that of the interfering cores?}

To explore how \ms captures microarchitectural effects arising from the mixed use of two memories with comparable MLP but significantly different latencies, we consider two experiments: (1) The core under observation targets \dr, while interfering cores target \pld. This case is labeled as "Obs: \dr, Int: \pld{}" and color-coded in red. (2) The core under observation targets \pld, while interfering cores target \dr, labeled "Obs: \pld, Int: \dr," in blue. Figure~\ref{fig:dm-md-bwboxplot} reports the results.

In the "Obs: \dr, Int: \pld" case, \dr initially outperforms \pld in isolation. However, as more interfering cores are added, \pld exhibits more stable performance, with fewer fluctuations compared to \dr. The results suggest that the degradation is not due to direct contention over \dr---since bandwidth is measured for \dr and the interfering cores stress \pld. Conversely, it suggests a bottleneck elsewhere in the system. At the highest interference level (three interfering cores), the results clearly reflect that saturating the \pld causes large performance degradation for accesses to the \dr. 

This effect would be counterintuitive without the previously examined MLP and latency analyses. Indeed, even with comparable MLP values, the higher latency of \pld (under stress) can delay DRAM transactions when their paths overlap in shared bus elements, such as at the level of CCI (Figure ~\ref{fig:zcu}). The increased latency of pending \pld transactions causes them to occupy shared bus queue entries longer, thereby reducing availability for \dr{}-bound requests.

\begin{figure}

\centering
\includegraphics[width=\linewidth]{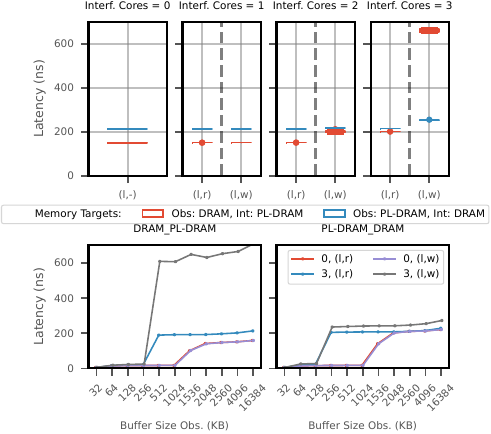}
\caption{Heterogeneous latency plots for DRAM and \pld under four stress scenarios with buffer size of 4 MB}
\label{fig:dram-mig-mig-dram-boxline}
\vspace{-0.5cm}
\end{figure}

A similar observation, but in reverse, is presented in Figure~\ref{fig:dram-mig-mig-dram-boxline}, where we focus on latency analysis. In the "Obs: \dr, Int: \pld" experiment, a noticeable increase in latency is observed when the heterogeneous system becomes congested. This indicates that \dr, despite its higher standalone bandwidth, is substantially more prone to latency degradation under high-stress mixed memory usage. The line plots (bottom of Figure~\ref{fig:dram-mig-mig-dram-boxline}) for the same case clearly highlight this trend. \pld is not affected by the issue, as shown by the results for the  "Obs: \pld, Int: \dr{}" case.

\subsection{Scratchpad Analysis} \label{sec:spmanalysis}

We showcase how \ms can be used to analyze the performance of scratchpad memories available in the system. These correspond to (1)~the On-Chip Memory (OCM) module on the PS side of the SoC and (2)~a Block RAM (BRAM) module on the PL side of the SoC---see Figure~\ref{fig:zcu}. 

In our platform, although the OCM capacity is 256 KB, only 128 KB is reserved for the memory pool, while the BRAM pool is 1 MB. Given the L1 and L2 cache sizes (32 KB and 1 MB, respectively), using cacheable operations would lead to cache hits, misrepresenting actual memory behavior.

Therefore, non-cacheable operations are necessary to conduct scratchpad memory analysis. 

To address this, we leverage the non-cacheable version of \ms's bandwidth (\texttt{s} for reads, and \texttt{x} and \texttt{y} for writes) and latency (\texttt{m}) workloads, as reported in Table~\ref{tab:accesspattern}.

As described in Appendix~\ref{sec:impl}, non-cacheable read workloads (\texttt{s} and \texttt{m}) perform a combination of cache line accesses followed by cache clean+invalidations. We always perform $500$ iterations in each scenario. Thus, after the first access and invalidation, all the subsequent accesses are ensured to miss in cache. We use two types of non-cacheable write operations. The first, denoted as \texttt{x}, issues store operations followed by cache invalidations. Since the cache policy is WAWB, reads from memory to load cache lines will still occur. Conversely, the \texttt{y} access strategy employs streaming writes, which follow a write-no-allocate policy, bypassing the cache. 

\subsubsection{\textbf{Homogeneous Bandwidth Analysis}}\label{sec:ocm-bram-bw}

\begin{figure*}

\centering
\includegraphics{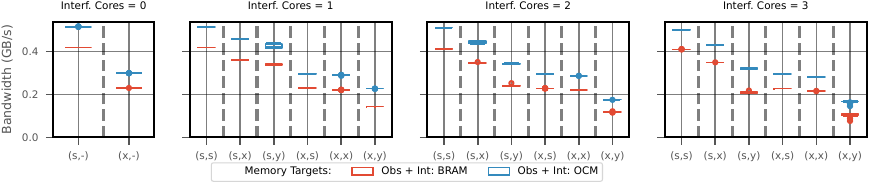}
\vspace{-0.2cm}
\caption{Homogeneous bandwidth results for OCM and BRAM under four stress scenarios with 32KB buffer.}
\label{fig:bram-ocm-boxplot}
\vspace{-0.5cm}
\end{figure*}

Figure~\ref{fig:bram-ocm-boxplot} presents our measurements for homogeneous bandwidth analysis of OCM and BRAM. 
As interference increases, OCM bandwidth progressively degrades from the \mpat{s,s} case to the \mpat{x,y} case. This trend mirrors the behavior discussed in Section~\ref{sec:dramanalysis}, where read operations are more vulnerable to interference due to the non-blocking nature of writes. When measuring read bandwidth, \mpat{s,y}, which employs write streaming, results in the lowest observed bandwidth. 
BRAM exhibits the same decreasing trend observed for OCM, and its absolute bandwidth remains consistently lower than OCM.

\begin{figure}

\centering
\includegraphics{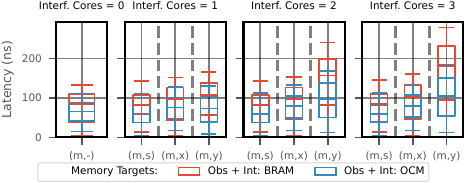}
\vspace{-0.2cm}
\caption{Homogeneous latency results for OCM and BRAM under four stress scenarios with 32KB buffer.}
\label{fig:bram-m-ocm-m-boxplot}
\vspace{-0.3cm}
\end{figure}

\subsubsection{\textbf{Homogeneous Latency Analysis}}
Figure~\ref{fig:bram-m-ocm-m-boxplot} reports latency results for BRAM (red) and OCM (blue) using the \texttt{m} access pattern.
With no interference, \emph{(m,-)} case, OCM outperforms BRAM, showing lower and more stable latency.

As interference increases, OCM maintains tighter and lower latency. In contrast, BRAM exhibits higher median latency across most interference, especially under \mpat{m,x} and \mpat{m,y}.
In conclusion, BRAM shows higher sensitivity to interference compared to OCM, making OCM a more reliable choice for the allocation of latency-critical memory pages.

\subsection{Cache Analysis} \label{sec:cacheanalysis}

In the experiments presented in this section, \rev{we leverage \ms to reproduce the effect of \emph{cache bank contention under hits} previously observed in~\cite{bank_cont_RTAS23}}. \rev{We also use this experiment to validate that the measurements obtained through \ms match those} observable with traditional benchmarks. In particular, we compare \ms to the \texttt{bandwidth} benchmark from the IsolBench suite\footnote{\url{https://github.com/CSL-KU/IsolBench/blob/master/bench/bandwidth.c}}.
First, Figure~\ref{fig:dram-isolbench-boxplot} compares two \dr bandwidth measurement experiments: one using IsolBench (color-coded in red), and the other using \ms (color-coded in blue). In both cases, the buffer size per core is set to 256\,KB: larger than L1 but small enough to fit within the LLC, ensuring that all accesses are hits.
\rev{Thus,} both experiments target the same memory module and follow equivalent configurations. \rev{The very close match in the measurements obtained using the two benchmarking approaches serves as validation that what is observed with \ms is indeed in line with established memory performance benchmark measurement toolkits, justifying further analysis relying solely on \ms.}

\subsubsection{\textbf{Bank contention under cache hits}}
\begin{figure}
\centering
\includegraphics[width=\linewidth]{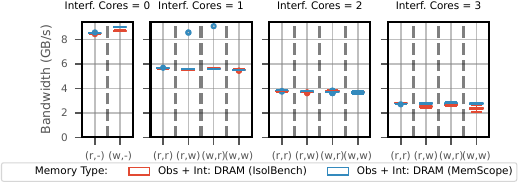}
\vspace{-0.5cm}
\caption{\rev{LLC bandwidth measurement for buffer size 256KB using the IsolBench suite (color-coded in red) and \ms (color-coded in blue).}}
\label{fig:dram-isolbench-boxplot}
\vspace{-0.5cm}
\end{figure}

Having ascertained that the cache-hit performance drop under stress identified by \ms is repeatable, we \rev{conduct a further experiment to verify that indeed the source of the performance degradation observed in Figure~\ref{fig:dram-isolbench-boxplot} can be attributed to the problem of cache bank contention, as} previously studied in~\cite{bank_cont_RTAS23}.

\begin{table}[t]
  \caption{Event counts under varying interference levels}
  \label{tab:event-interference}
  \centering
  {\scriptsize
  \begin{tabular}{lcccc}
    \toprule
    \textbf{Event/Interf. cores} & \textbf{Zero} & \textbf{One} & \textbf{Two} & \textbf{Three} \\
    \midrule
    \texttt{CPU\_CYCLE} & 17,131,051 & 26,228,725 & 39,834,512 & 53,836,500 \\
    \texttt{MEM\_ACCESS} & 2,049,051 & 3,764,331 & 3,760,759 & 3,748,782 \\
    \texttt{L2D\_CACHE} & 3,855,710 & 3,764,331 & 3,760,759 & 3,748,782 \\
    \texttt{L2D\_CACHE\_REFILL} & 5,182 & 204 & 1,748 & 5,591 \\ \hline
    Cache Hit Rate & 99.87\% & 99.99\% & 99.95\% & 99.85\% \\
    Cycles/Access & 4.44 & 6.97 & 10.59 & 14.36\\
    \bottomrule
  \end{tabular}
  }
\end{table}

To this end, we leveraged the integrated support for performance counter sampling in \ms. We sampled the counters on the core under observation, focusing on four key metrics listed in Table~\ref{tab:event-interference}: CPU cycles (\texttt{CPU\_CYCLE}), data memory accesses (\texttt{MEM\_ACCESS}), L2 data cache accesses (\texttt{L2D\_CACHE}), and L2 data cache refills (\texttt{L2D\_CACHE\_REFILL}). The results confirmed our hypothesis: cache hit rates ($>$99.8\%) aligned with expectations. However, the number of CPU cycles per cache access increased notably ($3.23\times$). As such, we conclude that the effect arises from cache bank-level contention on the hit path.

\begin{figure}

\centering
\includegraphics[width=\linewidth]{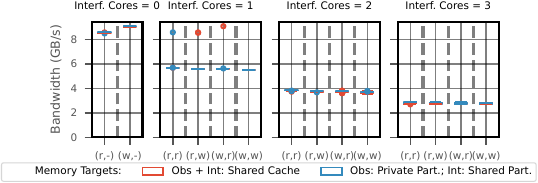}
\vspace{-0.4cm}
\caption{Bandwidth measurement, for buffer size 256\,KB, with and without cache partitioning.}
\label{fig:dram-all256-boxplot}
\vspace{-0.5cm}
\end{figure}

\subsubsection{\textbf{Bank contention under hits, with cache partitioning}}\label{sec:bankcont_jh}

Since cache bank contention on the \rev{hit} path~\cite{bank_cont_RTAS23} is unaffected by cache partitioning, we postulate that the same \rev{results should be obtainable with \ms}. Indeed, while cache partitioning divides the cache space, it does not deconflict the banks, so contention at the bank level remains. We \rev{use \ms to reproduce this effect for the first time on a platform featuring in-order Cortex-A53 cores with a single-bank LLC, while it was previously observed on out-of-order Cortex-A72, Cortex-A52, and Xuantie C910~\cite{bank_cont_RTAS23, cache_bank_regul_rtas24}}.

First, as mentioned in Section~\ref{sec:eval_meth}, we leverage the Jailhouse partitioning hypervisor to export a 25\% private L2 cache reservation as a memory pool, namely \texttt{pvtpool}. Since this is yet another pool, we can utilize the full array of benchmarks available in \ms. For this experiment, we focus on bandwidth behavior.

Next, we use a similar setup as per Figure~\ref{fig:dram-isolbench-boxplot} in Figure~\ref{fig:dram-all256-boxplot}, i.e., where all the cores hit in L2 cache (256\,KB buffers). However, in these experiments, we consider the cases in which partitioning is disabled (red) vs. enabled (blue). In the latter case, the observed core strictly allocates from the private cache partition (\texttt{pvtpool}). As expected, due to hit-path bank contention, partitioning is ineffective in mitigating contention.

\begin{figure}
\centering
\includegraphics{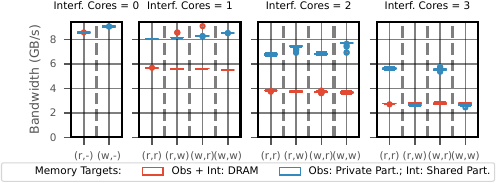}
\vspace{-0.2cm}
\caption{Bandwidth measurement with all cores accessing 256\,KB in shared cache partition (red) vs. observed core accessing 256\,KB from private cache partition and all interfering cores accessing 4\,MB from shared partition (blue).}
\label{fig:dram-isol-dram-jh-isol-boxplot}
\vspace{-0.3cm}
\end{figure}

\begin{figure}[t]
\centering
\includegraphics{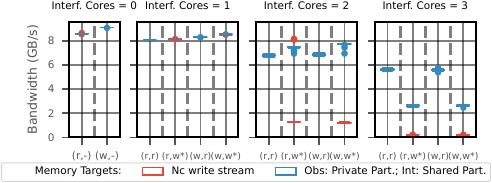}
\vspace{-0.2cm}
\caption{Bandwidth measurement for core under observation always accessing 256\,KB in private cache partition. Interfering cores access 4\,MB from shared cache partition w/ normal reads/writes (blue) vs. write-streaming (red).}
\label{fig:nczva}
\vspace{-0.5cm}
\end{figure}

\subsubsection{\textbf{Bank contention under miss, with cache partitioning}}

We conducted additional experiments to understand the conditions under which the system benefits from cache partitioning. Figure~\ref{fig:dram-isol-dram-jh-isol-boxplot} presents such a case. In these experiments, we consider two cases. In the first case (red), all the cores access 256\,KB from the shared cache partition---which is 3/4 of the L2, thus 768\,KB. Thus, the observed core suffers inter-core evictions. In the second case (blue), the observed core accesses 256\,KB mapping to the private cache partition (\texttt{pvtpool}), hitting in cache; all the stressor cores access 4\,MB from the shared cache partition, missing in cache. The plot shows that cache partitioning is effective for most types of interfering workloads. The only exceptions are the \mpat{r,w} and \mpat{w,w} cases, where miss-path cache bank contention occurs.

\ms allows us to push the effects of miss-path cache bank contention to the limit. To test this, in Figure ~\ref{fig:nczva}, we run an experiment where the core under observation always accesses 256\,KB from the private cache partition
while interfering cores use normal reads/writes (blue) to access 4\,MB from the shared cache partition---this is identical to the blue case in Figure~\ref{fig:dram-isol-dram-jh-isol-boxplot}. Next, we compare it to the case (red) where no changes are made to the observed core, while the interfering cores still access 4\,MB from the shared cache partition, but they do so using non-cacheable write-streaming operations---\texttt{y} access strategy, see Table~\ref{tab:accesspattern}. For the non-cacheable write-stream experiment, we evaluated the \mpat{r,y} and \mpat{w,y} combinations. Since some experiments included both \texttt{w} (normal cacheable write) and \texttt{y} (non-cacheable write stream) operations, we use the \texttt{w$^*$} notation in the plot. This corresponds to \texttt{w} for the case where stressors use normal writes (blue) and to \texttt{y} otherwise (red).

The results clearly show that while the measured bandwidth is identical in the case of one stressor core, drastic performance degradation is caused by streaming writes with two or more active interfering cores, in spite of cache partitioning. The very high performance loss (about $40\times$) is in line with similar results from~\rev{\cite{rtas2019_denial} also obtained on Cortex-A53 platforms.}

\subsection{Management of Real-Time Applications using \ms} \label{sec:realtime}

In this section, we investigate how applications can practically leverage the insights captured by \ms. 

Typically, real-time management of applications accessing shared memory subsystems relies on memory bandwidth regulation. In this subsection, we present a new dimension of management enabled by the insights captured through \ms and its ability to characterize the full heterogeneous memory subsystem. In addition to traditional bandwidth regulation, it becomes possible to make informed decisions about which memory type should be used by a given application. Understanding the memory characteristics and predicting their behavior under stress can significantly aid this determination.

\begin{figure}
\centering
\includegraphics[width=\linewidth]{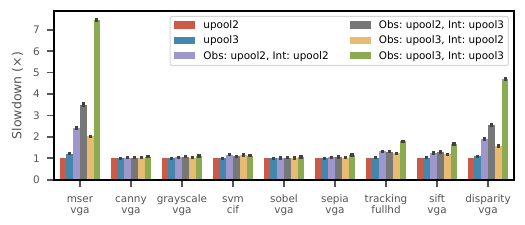}
\vspace{-0.8cm}
\caption{Heterogeneous memory management in real-world applications.}
\label{fig:rt-bench-barplot}
\vspace{-0.5cm}
\end{figure}

We perform our analysis on benchmarks from the San Diego Vision Benchmark Suite (SD-VBS)~\cite{sdvbs} and the Image Filters from RT-Bench~\cite{rtbench}. RT-Bench provides the ability to map the benchmark's heap to any of the pools exported by \ms to user space \emph{(upool)}, as described in Section~\ref{sec:design}. We investigate how their end-to-end runtime varies by changing where the heap is mapped and present the results in Figure~\ref{fig:rt-bench-barplot}.

We focus on \emph{upool2} and \emph{upool3}, which correspond to \dr and \pld, respectively. The $x$-axis reports the name of the benchmark alongside its input size. The $y$-axis depicts the slowdown, with each job's duration normalized to the baseline (1st bar) defined as the case where the application runs in isolation and allocates solely from \emph{upool2} (normal DRAM). The 2nd bar corresponds to the in-isolation run with the heap allocated in \emph{upool3} (\pld{}).

In the 3rd, 4th, 5th, and 6th bars in each cluster, the legend reports the \emph{upool} used to allocate the heap of the observed application, while write-heavy interference from 3 stressors is introduced targeting~\emph{upool2} and~\emph{upool3}, as per legend. 

The performance macro-trends observed in Figure ~\ref{fig:rt-bench-barplot} align with the insights captured by \ms. Indeed, although it may initially seem counterintuitive without \ms's guidance, allocating pages of the target application from DRAM (via \emph{upool2}) while stressors target \pld (via \emph{upool3}) results in higher slowdowns compared to the inverse setup. This is true across all the benchmarks and especially noticeable in benchmarks like \texttt{mser} and \texttt{disparity}.

\definecolor{Gray}{gray}{0.85}
\begin{table}
\centering
\caption{Comparison with other memory benchmarking tools}
\label{tbl:related}
\begin{minipage}[b]{\linewidth}
\resizebox{\linewidth}{!}{ 
\begin{tabular}{l c c c c c c}
\toprule
\multirow{2}*{Prior Work} & Open & Multi & Heterogeneous &  Kernel & Performance & Supported \\
& Source & Core & Memory & Mode & Counters & Architectures \\
\midrule
Intel MLC~\cite{intelmlc} & \No & \Yes & \Yes &  \No & \Yes & x86 \\
Isolbench~\cite{isolbench} & \Yes & \Yes & \No & \No & \Yes & x86/Arm \\
Nanoench~\cite{nanobench} &  \Yes & \No & \No &   \Yes & \Yes & x86 \\
Heimdall~\cite{heimdall} &  \Yes & \Yes & \Yes & \Yes & \No & x86 \\
LENS~\cite{lens} &      \Yes & \Yes & \Yes & \Yes & \No & x86 \\
tinymembench~\cite{tinymembench} & \Yes & \No & \No & \No & \No & x86/Arm \\
\rowcolor{Gray} \textbf{\ms} & \textbf{\Yes} & \textbf{\Yes} & \textbf{\Yes} & \textbf{\Yes} & \textbf{\Yes} & \textbf{Arm} \\
\bottomrule
\end{tabular}
}
\end{minipage}
\vspace{-0.5cm}
\end{table}
\section{Related Work}\label{sec:relwork}

Performance characterization is crucial for any heterogeneous system.
When a system features resource heterogeneity (in compute and/or memory), it must continuously decide how to optimally utilize these diverse resources for varying compute demands.
Making such decisions is only possible with a thorough understanding of the performance characteristics of each individual resource.
As a result, performance characterization has been the focus of many studies~\cite{heimdall, demystifying, izraelevitz2019basic}.
In particular, the performance of memory subsystems has received significant attention~\cite{softmc,heimdall,izraelevitz2019basic,cachetlb,thomborson2000measuring, coleman2001automatic}.
Most of these studies, however, focus on either cache behavior~\cite{cachetlb,thomborson2000measuring, coleman2001automatic} or a single memory technology~\cite{softmc,heimdall,izraelevitz2019basic, isolbench}, often in general-purpose, high-performance settings.
Furthermore, the majority are implemented in user space, making them subject to the limitations outlined in Section~\ref{sec:motivation}.
In contrast, Memscope offers a precise, extensible, kernel-level, open-source framework specifically designed for heterogeneous memory systems in embedded real-time environments.
Table~\ref{tbl:related} shows a high-level comparison of \ms with closely related memory benchmarking tools. In the rest of this section, we survey works in the broader area of memory characterization.

\paragraph{Characterizing Caches}
Several prior studies have proposed microbenchmark techniques to determine cache hierarchy parameters, such as cache size, associativity, block size, and latency~\cite{cachetlb,thomborson2000measuring, coleman2001automatic, dongarra2004accurate, yotov2005automatic, yotov2005capacity, molka2009memory, abel2013measurement,maurice2015reverse}.
These studies are performed either to guide performance optimization~\cite{cachetlb,thomborson2000measuring, coleman2001automatic}, or for performing cache side-channel attacks~\cite{maurice2015reverse,liu2015last}.
Most of these work assume a constant penalty for accesses that miss the cache and thus need to go to a single-technology main memory.
While \ms's microbenchmakrs also often need to consider caches--mostly to bypass them and reach to the main memory, Memscope's goal is different. It provides a benchmarking framework to precisely characterize a heterogeneous memory system beyond just cache properties.

\paragraph{Characterizing Single Memory Technology}
Prior work also extensively studied performance properties of a single memory technology as the main memory.
DRAM is perhaps the most studied one~\cite{softmc, drambender,trr,drama,dramaddress}. 
SoftMC~\cite{softmc} offers an open-source FPGA-based benchmarking platform that can test DRAM memory modules through a DDR interface, by directly sending DDR commands to the modules and measuring the response time. 
More recently DRAM Bender~\cite{drambender} builds on top of SoftMC and provides users the ability to write DRAM-based tests in high-level programming languages such as python. 
There are also other benchmarking studies that try to determine undocumented DRAM properties such as the refresh mechanism~\cite{trr}, DRAM row buffer~\cite{drama}, and DRAM address to row mappings~\cite{barenghi2018software,drama, dramaddress}.
Similarly there many studies to understand low-level device-level characteristics of other memory technologies such as  NVM~\cite{lens, nvleak, izraelevitz2019basic}, HBM~\cite{gpuhbm, shuhai, PENG201857, hbmbw}, and PIM~\cite{PIM}.
In contrast to these studies, \ms does not target only one memory technology, but focuses on understanding the entire heterogeneous memory system, including how different memories affect each other.

\paragraph{Characterizing Heterogeneous Memory}
The majority of prior work on characterizing heterogeneous memory systems has focused on general-purpose, high-performance computing.
Modern high-performance multicore servers typically feature a non-uniform memory access (NUMA) design, where clusters of cores share a single memory controller, and nodes are interconnected via high-speed links.
In such systems, any core can access memory attached to the entire system, but with non-uniform latency, as the access time depends on the memory location relative to the requesting core.
This introduces challenges similar to those in heterogeneous memory systems.
Several studies~\cite{ENTEZARIMALEKI2020172, numa2} have characterized NUMA performance to optimize overall system efficiency.
The introduction of persistent memory modules (such as Intel's Optane) has added another layer of heterogeneity in high-performance computing, and prior work has explored their performance characteristics~\cite{dram-nvm, izraelevitz2019basic, lens, nvleak}.
More recently, CXL (Compute Express Link) has emerged as a cache-coherent interconnect built on top of PCIe, allowing systems to add memory modules to the CXL fabric, introducing yet another form of heterogeneity.
The performance of CXL-based memory has been the focus of several recent studies~\cite{heimdall, Sun_2023}.

\paragraph{Generic Benchmarking Frameworks}
Prior work has also proposed generic microbenchmarking tools that allow users to infer performance characteristics of user-provided code, usually through measuring performance counters.
Linux perf~\cite{perf} allows users to measure performance counters for a particular executable.  Agner tool~\cite{agner} gives users more control by allowing measurement for a particular part of the code. Similarly, Nanobench~\cite{nanobench} also allows users to read performance counters of a microbenchmark written for x86 and runs in kernel mode. However, unlike \ms, nanobench does not provide multi-core microbenchmarks for heterogeneous memory characterization and only supports x86.

\section{Conclusion}\ms is a novel kernel-level memory benchmarking framework designed and implemented to characterize the temporal behavior of heterogeneous memory subsystems, particularly in real-time embedded systems. It is implemented entirely in kernel space to leverage privileged access to kernel APIs. This enables fine-grained control over core execution, physical memory allocation, and cache states.
\ms includes an extensible benchmark library not only for measuring bandwidth and latency but also for observing the relevant microarchitectural behaviors and events. 
Using \ms, we \rev{reproduce known effects and} provide several new insights into the performance behavior of an embedded system with heterogeneous memory. In addition, the insight offered by \ms can be leveraged to make counter-intuitive yet beneficial memory management decisions for real-time tasks to reduce their sensitivity to contention effects.

\section*{Acknowledgments}

Different co-authors used Grammarly and Chat-GPT only with the intent to assist in grammatical correction and enhancement. 

\bibliographystyle{ieeetr}
\bibliography{paper}

\newpage
\pagebreak
~~
\pagebreak
\appendix

In this appendix, we provide additional implementation details to ensure reproducibility. We also provide additional experimental results that were deemed comparably less interesting and thus not included in Section~\ref{sec:evaluation}.

\appendices

\section{Implementation Details}\label{sec:impl}

In this section, we present an overview of the key details regarding a proof-of-concept Linux implementation of the proposed \ms benchmarking infrastructure. \ms is currently implemented as a Linux 5.4 kernel module and does not require kernel modifications. The proposed implementation primarily targets ARMv8 architectures, which largely dominate the landscape of high-performance embedded systems. 

\subsection{Memory Pool Manager Implementation}
Given our focus on heterogeneous embedded systems, \ms is designed to detect the available memory modules automatically. It does so by leveraging existing kernel infrastructure to describe hardware modules, namely Device Tree Blobs (DTBs). In SoCs that do not support (or only partially support) hardware enumeration (e.g., PCIe), DTBs are provided to the kernel at boot time by the bootloader. A DTB contains a description of the hardware components of the system and is utilized by the Linux kernel for initialization purposes. A DTB can be generated by assembling one or more Device Tree Source (DTS) files using the Device Tree Compiler (DTC). 

The pool manager in \ms looks for any memory node in the boot-time DTB with the \texttt{"mempool"} value for the \texttt{compatible} property, as depicted in the corresponding DTS reported in Figure~\ref{fig:dts}. For each discovered node, \ms retrieves the start address and size by reading the \texttt{reg} property using the \texttt{of\_property\_read\_u64\_index} function. Indeed, the first two values of the \texttt{reg} property encode a 64-bit start address value (e.g., \texttt{0x0a0000000} for the BRAM pool defined in Figure~\ref{fig:dts}), while the latter two values encode a 64-bit aperture size in bytes (e.g., \texttt{0x000100000} = 1MB for the BRAM pool defined in Figure~\ref{fig:dts}). This information is used to create the corresponding allocation pool.

To initialize a memory pool, the memory pool manager first maps the corresponding memory aperture into kernel memory using \texttt{memremap}. The resulting kernel virtual address (KVA) is used for the next step. Here, the manager leverages the \texttt{genpool}\footnote{See official documentation at \url{https://www.kernel.org/doc/html/v4.17/core-api/genalloc.html}.} Linux kernel API to create (\texttt{gen\_pool\_create}) an ad-hoc allocation pool, populating it (\texttt{gen\_pool\_add}) with all the pages in the previously obtained KVA range. Upon initialization, each pool is assigned a unique ID which can later be used to construct experiments targeting individual pools.
Upon removal of the kernel module, the memory pool manager destroys the pool (\texttt{gen\_pool\_destroy}) and performs any necessary clean-up operations.

\begin{figure}
\lstset{
    language=C,
    basicstyle=\ttfamily\footnotesize, 
    frame=single,
    breaklines=true
}
\begin{lstlisting}
bram@a0000000 {
    device_type = "memory";
    compatible = "mempool";
    reg = <0x0 0xa0000000 0x0 0x100000>;
};

dram@10000000 {
    device_type = "memory";
    compatible = "mempool";
    reg = <0x0 0x10000000 0x0 0x10000000>;
};
\end{lstlisting}
\caption{Device Tree Source (DTS) for \ms-compatible memory nodes.}
\label{fig:dts}
\end{figure}

\subsection{Workload Library Implementation}

\textbf{Configurable Buffer Initialization:} Buffer allocations from the selected pool are performed via the \texttt{gen\_pool\_alloc} API. The buffer initialization depends on the type of workload. For bandwidth test benches, buffers are filled sequentially with integer values. This is only useful for sanity checking that no buffer corruption has occurred, e.g., after introducing a new type of experiment. 

Buffer initialization for latency measurements is more complex. In the latter case, the goal is to force data dependencies to minimize the number of outstanding memory transactions. Thus, the buffer is initialized with a chain of indices: the first cache line holds the index to the next cache line, and so on. The structure of the dereference chain is randomized to ensure that no prefetching occurs, while ensuring that the chain spans the entire size of the buffer with no repeated accesses. Figure~\ref{fig:lat_shuffle} provides an intuitive description of the latency buffer initialization strategy. Initially (Step 1), the buffer is initialized with a sequential chain of references, one per cacheline. Next, (Step 2) a permutation array \texttt{perm} is created via a series of $k$ subsequent swaps. Finally, (Step 3) the original buffer is updated by following the permutation buffer. Specifically, the pointer in cacheline \texttt{perm[}$i$\texttt{]} is updated to point to the cacheline with index \texttt{perm[}$i+1$\texttt{]}.

\begin{figure*}
    \centering
    \input{figures/latency_init}
    \caption{Initialization of latency buffer for random but full walk over data-dependency buffer.}
    \label{fig:lat_shuffle}
\end{figure*}

\noindent\textbf{Test Bench Algorithm and Structure:}
 \ms implements five low-level functions that correspond to the various access types supported for bandwidth benchmarking, detailed as follows:  
\texttt{\_\_access\_bw\_read} and \texttt{\_\_access\_bw\_write} use \texttt{ldr} and \texttt{str} assembly instructions with post-increment for efficiency. Similarly, for reading/writing bandwidth measurements using non-temporal instructions, we use \texttt{ldnp} and \texttt{stnp} instructions.

Non-cacheable read operations for bandwidth measurement are implemented in two different ways. The first implementation, called \texttt{\_\_NC\_IMPL\_DCAFTER}, loads the address from memory using \texttt{ldr} with post increment addressing. Then, this incremented address is immediately invalidated from the cache using the \texttt{dc civac} instruction. The only drawback is that the access to the very first cacheline in the buffer at each iteration might result in a hit.

The second approach, \texttt{\_\_NC\_IMPL\_DCADD}, addresses the limitation of the first method. This approach first loads the address, then cleans and invalidates the same address, mitigating the first-access cache hit issue. The address is manually incremented to the following location using the \texttt{add} instruction.

We implemented two types of non-cacheable write-based operations. The first type is a store-based operation, implemented similarly to \texttt{\_\_NC\_IMPL\_DCAFTER}, but using the \texttt{str} (store) instruction. The second type is non-cacheable write stream, we use the special ARM AArch64 assembly instruction \texttt{dc zva}. This instruction writes a cacheline size of zero to the memory, skipping the cache allocation, and the rest of the loop is implemented as \texttt{\_\_NC\_IMPL\_DCADD}. 

Latency functions are implemented only in the read access pattern, where each element of the initialized buffer from the previous step is accessed. The loop continues until the pointer we are looking at is the same as the one we started from. For non-cacheable latency, after each access, the next address is invalidated, as in \texttt{\_\_NC\_IMPL\_DCAFTER}.

\noindent\textbf{Performance Counter Implementation:}
The performance counter implementation is straightforward. In ARM Cortex-A53, which is the platform for our experiments, performance counters are accessed through the Performance Monitoring Unit (PMU). Firstly, we enable the Performance Monitor Control Register (PMCR). To utilize the counters, specific bits must be set in this control register. Specifically, the \texttt{C} and \texttt{P} bits in the PMCR register need to be configured. The \texttt{C} bit (Clear) is used to reset the counters, ensuring a fresh start for measurements. The \texttt{P} bit enables counting of performance events. After enabling the performance counters, we must configure the specific counters to be used, taking into account the limitation of six counters per core.
We achieve this by setting the corresponding bits in the \texttt{pmcntenset\_el0} (Performance Monitor Counters Enable Set register). Finally, we need to write the event ID number we want to sample for, in \texttt{pmevtyperX\_el0} register (Performance Monitor Event Type Register X). It should be noted that ARMv8 provides multiple \texttt{pmevtyper} registers for performance monitoring. Finally sampling happens by reading the value of the performance monitoring counter X (pmevcntrX\_el0).

\subsection{Core Coordination Implementation}
The experiment validator checks parameters using simple if-else Statements and terminates the process if validation fails. Next, based on the selected test benches and the scenario to be executed, the workload buffer allocation and initialization are performed, as previously described. 

\noindent\textbf{Remote Core Scheduling:}Launching remote activities is implemented using the \texttt{on\_each\_cpu\_mask()} Linux kernel API, which allows specifying a function to be executed on each CPU of the system. To selectively run functions, such as activity stress or idle, on specific CPUs, we utilize the \texttt{cpumask} to set the desired CPUs. Each time \texttt{on\_each\_cpu\_mask()} is invoked, we define the function and the target core. This enables us to schedule specific functions on designated tasks.
Before setting the mask, we clear it using \texttt{cpumask\_clear}, and then loop over the CPUs, setting the mask  with regard to the current scenario, for each core using \texttt{cpumask\_set\_cpu()}. This requires preparing the correct mask for core selection, ensuring the appropriate activity is assigned to the correct core.
We prepare two sets of masks: one for the cores to execute the activity stress and one for those to remain idle. Since this process is performed in the main loop, the masks are adjusted based on each scenario. For each scenario, we loop over the CPUs and assign cores to specific activities, ensuring that the local observation core is not assigned to any other activity. We call \texttt{on\_each\_cpu()} twice: once for the activity stress function and once for the activity idle function, executing them consecutively with the corresponding masks.

\noindent \textbf{Measurement Coordination Mechanism:}We use spinlocks to implement our notification system for measurement coordination in an unconventional manner. Each core has its own lock, which is initialized before the main activity loop using \texttt{spin\_lock\_init()}. When the core under observation is waiting for remote cores to start or stop their execution, it spins on their locks, continuously checking whether their locks are on or off.
The core under observation starts spinning and waits for all locks to be acquired using \texttt{spin\_lock()}. When all the locks are acquired, it indicates that all remote cores have started their respective remote activities. At this point, the main activity can begin and the measurement either time or performance counter samples start.
Once the main activity concludes, and the core coordinator instructs remote cores to stop, the core under observation spins on the locks again to ensure all locks are released, signifying the remote executions are complete. This allows the core under observation to proceed to the next scenario. Thus, the main activity is "sandwiched" between two phases of \texttt{spin\_lock} spinning, ensuring that measurements are taken at the correct times.

\noindent We define a global variable \texttt{g\_exp\_running} to control the start and stop of remote execution. To acquire timing samples, we use the Linux kernel function \texttt{ktime\_get\_ns()}, which provides precise time measurements in nanoseconds as part of the \texttt{ktime} API. All measurements—both time and performance counters, if applicable—are taken just before starting the main activity, after the core coordinator ensures that all remote activities have started, and exactly when the main activity finishes.
To collect performance counter samples, we read the register \texttt{pmevcntri\_el0}, which depends on the counter number being used. The difference between these two samples provides the desired measurement. To ensure the most accurate measurement possible, we disable interrupts and preemption using \texttt{local\_irq\_save}. Once the measurement is complete, we restore the normal status with \texttt{local\_irq\_restore(flags)}. Additionally, to prevent CPU migration, we pin each activity to its assigned core using \texttt{put\_core}, and restore the original core assignment once the experiment concludes using \texttt{get\_cpu()}. When the experiment is over, results are collected, and either bandwidth or latency, depending on the workload, is calculated and sent to the user interface. The final phase is the clean-up, where all allocated buffers are 
freed using the \texttt{gen\_pool\_free} function.

\noindent It should be noted that, since we have the potential to run cacheable and non-cacheable operations consecutively, we clean and invalidate the cache before starting a new scenario to ensure no targeted addresses from the cacheable experiment remain in the cache. This procedure is implemented mainly by using these instructions: reading content of counter timer register \texttt{ctr\_el0 } using \texttt{mrs} instruction and extracts bit 16 -19 using \texttt{ubfm}. After aligning the start address to the cache line boundary, we clean and invalidate each cache line in the loop using \texttt{dc civac} instruction.

\subsection{User Interface Implementation}\label{sec:uiimp}

The user interface kernel module is integrated into the \ms main module. Upon insertion, it establishes a communication channel between user and kernel space using \texttt{debugfs}, a virtual file system mounted in \texttt{sysfs}, providing debugging information and exposure to the kernel data structures.

During the initialization phase, the user interface module configures the necessary \texttt{debugfs} entries to enable communication with the kernel module.
First, \texttt{debugfs\_create\_dir} creates a directory named \texttt{membench} in the \texttt{debugfs} file system. If successful, it returns a pointer to the \texttt{dentry} structure of the directory. The \texttt{dentry} structure carries the file path name, along with other useful information for file management in the kernel file system.

We have five main entries in our \texttt{debugfs} directory: \texttt{experiment}, \texttt{pools}, \texttt{cmd}, \texttt{perfcount}, and \texttt{results}. Each entry is implemented as a file and has its own set of file operations. These entries are created using \texttt{debugfs\_create\_file} with the appropriate permissions and file operations based on their configuration.
\texttt{experiment} has permission \texttt{0644}, meaning it is readable and writable by the owner (root) and readable by others. It supports both \texttt{read} and \texttt{write} file operations. In read mode, it provides information about the most recent experiment conducted, as interpreted by the kernel module. When written, it allows users to define a new benchmarking experiment setup. The user data is read using the \texttt{copy\_from\_user} function, which copies it to the destination buffer in the kernel memory space. This kernel buffer is then processed by \texttt{sscanf}, a standard C library function, which reads the data in a formatted way to populate the internal data structures for the experiment parameters.
\texttt{pools} has the permission \texttt{0444}, which means it is readable by everyone (owner, group, and others) but not writable. It provides a read-only listing of the detected memory pools and their initialization status. \texttt{results} shares the same permission and operational mode as \texttt{pools}. When read, it displays the result information using \texttt{seq\_printf}.
\texttt{perfcount} and \texttt{cmd} both have read and write operations with the permission \texttt{0644}. When written to, they receive user data---event numbers for \texttt{perfcount} and commands for \texttt{cmd}---using \texttt{copy\_from\_user}. In read mode, they display the performance counters setup and the chosen commands, respectively.
Upon disabling the module and removing its kernel module, \texttt{debugfs\_remove\_recursive()} is called to recursively remove all the contents of the \texttt{membench} directory from \texttt{debugfs} and clean up.

In addition to the aforementioned responsibilities, the user interface component is extended to expose each kernel's internal memory pool, created by the memory pool manager, to the user space. For implementing this part, we use another virtual file system in the Linux kernel---the \texttt{/dev} filesystem. Files created under \texttt{devfs} filesystem represent virtual devices and serve as an interface for user space to interact with kernel-space components representing drivers and hardware devices.
Similar to \texttt{debugfs}, upon kernel module insertion, the user-exposed memory pools are initialized under the path \texttt{/dev/upool<ID>}, where \texttt{ID} denotes the memory pool index. This initialization includes the automatic creation of \emph{upool} device nodes under \texttt{/dev/upool<ID>}, in which each \emph{upool} node is implemented as a file descriptor with the following main file operations: \emph{open}, \emph{release}, and \emph{mmap}.
The main API used in the driver's open handler is \texttt{iminor()}, which allows the kernel to identify which \emph{upool} is being accessed. The release handler prepares the \emph{upool} instances for closure by retrieving the relevant information associated with the target \emph{upool} and deallocating any previously mmap'd memory range.
The main file operation for \emph{upool} file descriptors is the \emph{mmap} function. Whenever a user application invokes \texttt{mmap} on a given \texttt{/dev/upool<ID>} file, the requested pages are allocated from the corresponding \ms memory pool.
\emph{mmap} uses texttt{gen$\_$pool$\_$alloc} function from \texttt{genpool} API, which, if successful, returns a valid kernel virtual address as the beginning address of the memory buffer. Then, using \texttt{virt\_to\_page()} and \texttt{page\_to\_pfn()} respectively, the kernel virtual address is first converted to its corresponding page descriptor, and then the physical page's index---i.e., the page frame number (PFN)---is retrieved. Finally, \texttt{remap\_pfn\_range()} maps the physical page range into user-space memory. This enables the user application to access physical memory through the \texttt{mmap()} system call.
Similar to \texttt{debugfs}-based interfaces, removing the kernel module destroys the \emph{upools}, deallocating the associated memory pages and cleaning up the associated data structures.

\section{Additional Results on Management of Real-Time Applications using \ms}

\begin{figure}[h]
\centering
\includegraphics[width=\linewidth]{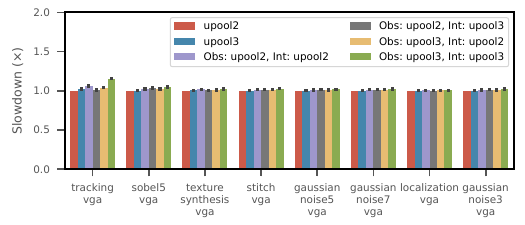}
\caption{Heterogeneous memory management in more real-world applications.}
\label{fig:rt-bench-barplot-extra-appendix}
\end{figure}

We present additional experiments related to Section~\ref{sec:realtime}. These results did not exhibit interesting trends, so we excluded them from Figure~\ref{fig:rt-bench-barplot} due to space constraints. The experimental setup for the results shown in Figure~\ref{fig:rt-bench-barplot-extra-appendix} is identical to the setup used for the experiments in Figure~\ref{fig:rt-bench-barplot}.

\end{document}